\documentclass[11pt, letterpaper]{article}

\usepackage{graphicx}
\usepackage{amsmath}
\usepackage{amssymb}
\usepackage{setspace}
\usepackage{epstopdf}
\usepackage{color}
\usepackage[letterpaper,left=1in,right=1in,top=1in,bottom=1in]{geometry}
\usepackage[linesnumbered,ruled,vlined,longend]{algorithm2e}
\usepackage{multirow}
\usepackage{longtable}
\usepackage{rotating}
\usepackage{threeparttable}
\usepackage{colortbl}
\usepackage{amsthm}
\usepackage[modulo]{lineno}
\usepackage{enumerate}

\newtheorem{remark}{Remark}

\newcommand{\m}{\mathbb}

\title{\Large \bf Robust economic MPC of the absorption column in post-combustion carbon capture through zone tracking}

\author{\centerline{\normalsize Benjamin Decardi-Nelson$^{a}$, Jinfeng Liu$^{a,}$\thanks{Corresponding author: J. Liu. Tel: +1-780-492-1317. Fax: +1-780-492-2881. Email: jinfeng@ualberta.ca}}\vspace{5mm}\\
    \centerline{\small $^{a}$ Department of Chemical \& Materials Engineering, University of Alberta,}\\
    \centerline{\small Edmonton, AB, Canada, T6G 1H9}}

\allowdisplaybreaks

\begin{document}

\date{}

\maketitle
\setstretch{1.39}

\begin{abstract} 

Several studies have reported the importance of optimally operating the absorption column in a post-combustion CO$_2$ capture (PCC) plant. It has been demonstrated in our previous work how economic Model Predictive Control (EMPC) has a great potential to improve the operation of the PCC plant. However, the use of a general economic objective such as maximizing the absorption efficiency of the column can cause EMPC to drive the state of the system close to the constraints. This may often than not lead to solvent overcirculation and flooding which are undesirable. In this work, we present an EMPC with zone tracking algorithm as an effective means to address this problem. The proposed control algorithm incorporates a zone tracking objective and an economic objective to form a multi-objective optimal control problem.  To ensure that the zone tracking objective is achieved in the presence of model uncertainties and time-varying flue gas flow rate, we propose a method to modify the original target zone with a control invariant set. The zone modification method combines both ellipsoidal control invariant set techniques and a back-off strategy. The use of ellipsoidal control invariant sets ensure that the method is applicable to large scale systems such as the absorption column. We present several simulation case studies that demonstrate the effectiveness and applicability of the proposed control algorithm to the absorption column in a post-combustion CO$_2$ capture plant. 


\end{abstract}


\section{Introduction}\label{sec:introduction}

Climate change is one of the most pressing global issues that needs immediate remedy to avoid catastrophic consequences on the future generations. A major contributing factor to climate change is the presence of large quantities of anthropogenic greenhouse gases especially carbon dioxide (CO$_2$) in the atmosphere. A major contributor to the increase in anthropogenic CO$_2$ in the atmosphere is due to the combustion of fossil fuels such as coal for electricity generation \cite{HOOK2013}. Several studies have shown that switching to low carbon energy sources such as renewable energy sources can help address the climate change issues \cite{BATAILLE2016}. However, the pursuit for relatively cheaper and reliable energy sources in response to the ever increasing demand for energy is making it difficult to switch to these lower carbon energy sources. In 2020, fossil fuels contributed to about 61\% of the global electricity generation sources with coal taking up roughly 35\% of the share \cite{BP2021}. Clearly, it is impractical to completely eliminate fossil fuels from the energy generation sources at a go. Therefore, effective means of reducing the emissions from fossil fuel power plants is the best approach to reduce CO$_2$ emissions while allowing the low carbon electricity generation technologies to reach maturity. Several approaches have been proposed to reduce CO$_2$ emissions from large point sources such as power plants. These approaches include pre-combustion, oxy-fuel combustion and post-combustion. However, amine-based post-combustion CO$_2$ capture (PCC) is the most mature and viable technology available today.

In amine-based PCC, the flue gas produced after combusting the fossil fuel is sent to a gas processing unit (PCC plant) where the amount of CO$_2$ in the gas is reduced using an amine solvent before being released into the atmosphere. This makes it easier to retrofit it into existing power plants. However, amine-based PCC is not without downsides. It has been shown that attaching a PCC plant to a power plant reduces the efficiency of the power plant by about 10\% for state-of-the-art monoethanolamine (MEA) solvent \cite{DAVISON2014}. This is because of the high energy requirements to regenerate the amine in the desorption unit. Sakwattanapong and coworkers \cite{SAKWATTANAPONG2005} demonstrated that maintaining the CO$_2$ concentration in the amine solvent within an optimal range during absorption is essential for efficient operation of the regeneration unit. It is therefore critical that advanced model-based process control techniques such as model predictive control (MPC) are employed in the control of the PCC plant.

Model Predictive Control is an advanced model-based optimal control method that has gained popularity within the chemical process industry. This is because of its ability to handle complex multivariable systems and constraints. Within the context of process control of PCC plants, several control schemes have been developed using MPC. Panahi and Skogestad \cite{PANAHI2012} investigated different control schemes using a linear MPC. He et al.  \cite{HE2016} also used a combined scheduling and MPC scheme to achieve the desired carbon dioxide absorption efficiency in the absorption column as well as the CO$_2$ purity in the gas outlet of the desorption unit. Bankole and coworkers  \cite{BANKOLE2018} investigated the flexibility of operating a PCC plant attached to a load following power plant using MPC. To address the presence of uncertainties in the control system, Patron and Ricardez-Sandoval implemented a robust MPC algorithm for the absorption column \cite{PATRON2020} as well as an integrated control and state estimation scheme for the PCC plant \cite{PATRON2022}. More recently, a variant of model predictive control (MPC) with a general objective known as economic MPC (EMPC) has received significant attention \cite{RAWLINGS2012,liu2016}. The objective function in an EMPC scheme generally reflects some economic performance criterion such as profit maximization or waste minimization. This is in contrast to the standard MPC where the objective is a positive definite quadratic function. The integration of process economics directly in the control layer makes EMPC of interest in many areas especially in the process industry. Decardi-Nelson, Liu and Liu \cite{DECARDI_NELSON2018} demonstrated the superiority of EMPC scheme over the standard MPC scheme in a full cycle PCC plant. 

While EMPC is a promising control algorithm for the PCC process, the general economic objective such as maximizing the absorption efficiency of the absorption column may drive the system states to the constraints. This may lead to column flooding and/or solvent overcirculation in the PCC plant. Column flooding can compromise the safety of both the absorption column and the personnel that manage it while solvent overcirculation may lead to high energy requirements for regeneration of the solvent in the desorption column. In another direction, it is not well understood how the presence of uncertainties affect the economic performance of EMPC in general. This is because of the integration of process economics in the control layer. Uncertainties are unavoidable in the real world. They are caused by the use of imperfect process models in the model-based control algorithms and/or unmeasured disturbances. A typical disturbances in a load following PCC plant attached to a power plant is large fluctuations in the flue gas flow rate \cite{RUA2020}. The presence of uncertainty in a control system can result in significant performance degradation and/or loss of stability.  This can ultimately compromise safety during operation. A common technique to address uncertainties in a control system is through robust MPC. However, as observed in a recent study by Patron and Ricardez-Sandoval \cite{PATRON2020}, the online computational requirements of robust MPC techniques such as the multi-scenario approach can be demanding as the number of uncertainties and scenarios increase. Moreover, for EMPC, it was pointed out in the study by Bayer and coworkers \cite{bayer2014} that simply transferring robust MPC techniques to EMPC could result in poor performance. This is because economic optimization and robustness are two objectives and often may conflict with each other. Robust MPC techniques have been designed to reject all disturbances to achieve their desired goal which may not be the case in EMPC as some disturbances can lead to better economic performance. It is therefore important to develop robust EMPC algorithms which does not involve complex online computations.

In this work, we present an economic MPC with zone tracking algorithm for the control of the absorption column of the PCC process under additive state uncertainties and time-varying flue gas flow rate. The proposed control algorithm only makes use of the nominal process model without explicitly accounting for the uncertainties in the process. This makes the online computations less demanding compared to the scenario-based approach. The integration of zone tracking in EMPC allows for concurrent handling of two objectives while enhancing the degree of robustness of the controller \cite{liu2019}. Zone MPC have been reported in several process control application areas such as diabetes management \cite{grosman2010}, control of building heating system \cite{privara2011}, control of irrigation systems \cite{mao2018} and control of coal-fired boiler-turbine generating system \cite{zhang2020}. This work builds on our previous EMPC with zone tracking formulation -- with \cite{DECARDI_NELSON2022} and without \cite{liu2018} uncertainty consideration. 
In line with the work by Decardi-Nelson and Liu \cite{DECARDI_NELSON2022}, we propose to track a control invariant subset of the target zone contrary to tracking the target zone. However, because of the large number of states in the process model of the absorption column, the zone modification algorithm developed by Decardi-Nelson and Liu \cite{DECARDI_NELSON2022} is computationally intractable. We therefore propose a target zone modification algorithm using ellipsoidal control invariant set computation techniques and a back-off strategy. This has a potential to extend the applicability of the robust EMPC with zone tracking algorithm to much wider range of systems.

The remainder of this paper is organized as follows. Section 2 presents the preliminaries which includes a description of the process model of the absorption column and the control problem to be solved. Section 3 describes the proposed robust economic MPC framework and the computation of the modified zone. In Section 4, extensive simulations are carried to demonstrate the efficacy of the proposed control algorithm. We summarize the key findings in this paper and discuss possible future directions in Section 5.

\section{Preliminaries}\label{sec:process_model}

\subsection{Notation} \label{subsec:notation}
Throughout this work, the symbol $\| \cdot \|_n$ denotes the $n$-norm of a scalar or a vector, the operator `$\backslash$' means set subtraction such that $\m A \backslash \m C = \{ x : x \in \m A, x \notin \m C  \}$, $\m R_+$ denotes the set of all real numbers greater than or equal to zero, the set $\m B$ represents the unit ball.

\subsection{Process description}\label{subsec:process_description}
\begin{figure}[htbp]
  \begin{center}
    \includegraphics[width=0.4\textwidth]{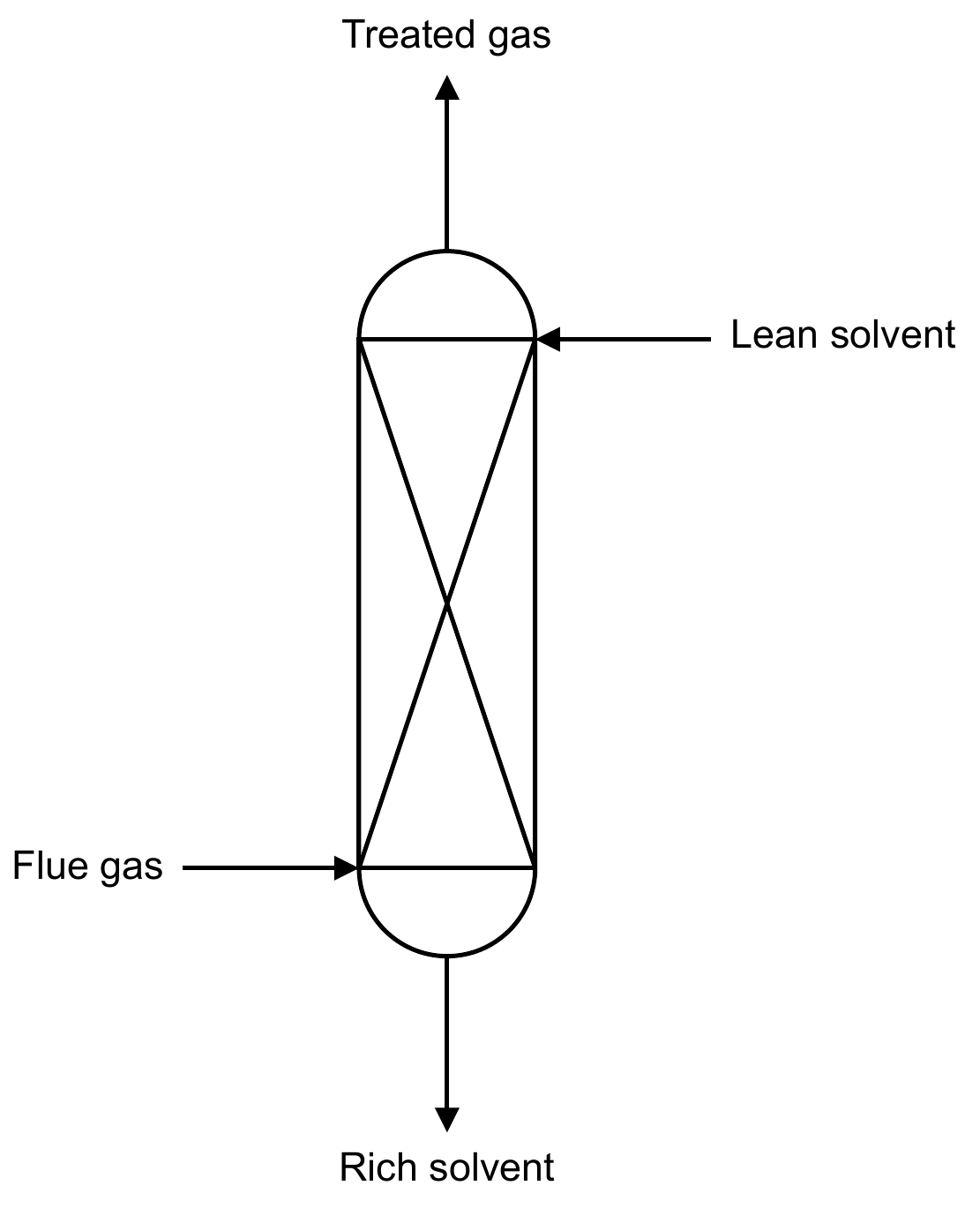}    
    \caption{A schematic diagram of a packed absorption column} 
    \label{fig:absorption_column}
  \end{center}
\end{figure}

An absorption column in an amine-based post-combustion CO$_2$ capture process is a multistage gas processing unit in which an amine solvent selectively removes CO$_2$ from the flue gas. The amine solvent with low amount of CO$_2$ (lean solvent) is introduced at the top of the column while the flue gas enters the column from the bottom in a counter-current manner as shown in Figure~\ref{fig:absorption_column}.
The absorption column is usually filled with packing materials to increase the contact area for mass transfer between the liquid and the gas phases. Following the transfer of CO$_2$ from the gas phase to the liquid phase, the liquid with increased amount of CO$_2$ (rich solvent) exits the column at the bottom while the treated gas exits the column at the top.

Owing to the reactive nature of the mass transfer process occurring in the absorption column, the rate-based approach is used to model the process. The rate-based model has been found to be superior to the equilibrium-based approach to modeling reactive mass transfer processes \cite{wang2011}. The following assumptions were used in modeling the absorption column:

\begin{itemize}
    \item The liquid and the gas phases are well mixed with no spatial variations in properties.
    \item The reactions are described using enhancement factor and occur only in the liquid phase.
    \item The heat and mass transfer occurring at the gas-liquid interface is described by the two film theory.
    \item The pressure drop along the axial direction of the column is linear.
    \item The velocities of the liquid and gas phases in the column remain constant.
    \item The absorption column is well insulated.
\end{itemize}

The simultaneous heat and mass transfer process occurring in the column is described by the partial differential equations in Equations~(\ref{eqn:liquid_mass_balance}) -- (\ref{eqn:gas_temperature_balance}).

\begin{equation}\label{eqn:liquid_mass_balance}
    \frac{\partial c_{Li}}{\partial t} = \frac{4F_L}{\pi D_c^2}\frac{\partial c_{Li}}{\partial z} + N_i a^I
\end{equation}

\begin{equation}\label{eqn:gas_mass_balance}
    \frac{\partial c_{Gi}}{\partial t} = -\frac{4F_G}{\pi D_c^2}\frac{\partial c_{Gi}}{\partial z} - N_i a^I
\end{equation}

\begin{equation}\label{eqn:liquid_temperature_balance}
    \frac{\partial T_{L}}{\partial t} = \frac{4F_L}{\pi D_c^2}\frac{\partial T_{L}}{\partial z} + \frac{Q_L a^I}{\sum_{i=1}^n c_{Li}c_{pi}}
\end{equation}

\begin{equation}\label{eqn:gas_temperature_balance}
    \frac{\partial T_{G}}{\partial t} = -\frac{4F_G}{\pi D_c^2}\frac{\partial T_{G}}{\partial z} + \frac{Q_G a^I}{\sum_{i=1}^n c_{Gi}c_{pi}}
\end{equation}
In Equations~(\ref{eqn:liquid_mass_balance}) -- (\ref{eqn:gas_temperature_balance}), $c_i$ represents the phase concentration of component $i$ in $kmol/m^3$, $F$ is the phase volumetric flow rate in $m^3/s$, $D_c$ is the diameter of the column in $m$, $N_i$ is the mass transfer rate of component $i$ in $kmol/m^2s$, $z$ is the height of the column in $m$, $T$ denotes the phase temperature in $K$, $Q$ denotes the heat transfer rate in $kJ/m^2s$, $c_p$ is the heat capacity in $kJ/kmol$, $a^I$ is the interfacial area in $m^2/m^3$. Also, subscripts $L$ and $G$ represent the liquid and gas phase respectively, and subscript $i$ denotes the components in the system namely CO$_2$, N$_2$, H$_2$O and MEA.

In the mathematical model above, the enhancement factor approach together with the Chilton-Colburn analogy \cite{CHILTON1934} is used to determine the influence of the reactions on the rate of heat and mass transfer of CO$_2$ from the gas phase to the liquid phase. Details of the physical and chemical properties of the components in the system can be found in the work by Decardi-Nelson et al. \cite{DECARDI_NELSON2018}. 

\subsection{Model discretization and state space representation}\label{subsec:model_discretization}
To avoid having to formulate and solve infinite dimensional optimal control problems online, the partial differential equations are converted to ordinary differential equations using the method of lines (MOL). The method of lines involves spatially discretizing the partial derivatives with the length of the column to obtain only differential equations with respect to time. In this work, the derivatives with respect to the length of the column were discretized into five stages to obtain 50 ordinary differential equations. Considering the presence of uncertainties in the system, the dynamics of the CO$_2$ absorption column can be written in a nonlinear state space model of the form:
\begin{equation}\label{eqn:discretized_model}
    \dot{x}(t) = f(x(t),u(t)) + w(t)
\end{equation}
where $\dot{x} \in \m{R}^{50}$ is the time derivative of the state, $x \in \m{R}^{50}$ is the state of the system at time $t \in \m R_+$, $u=F_L \in \m{R}$ is the manipulated input, and $w \in \m{R}^{50}$ represents the additive uncertainties that may be present in the system. 
The controlled output $y$ of the system is the CO$_2$ absorption efficiency and is given by
\begin{equation}\label{eqn:output}
    y(t) = h(x(t)) = \frac{\text{Molar flow rate of CO}_2 \text{ in} - \text{Molar flow rate of CO}_2 \text{ out}}{\text{Molar flow rate of CO}_2 \text{ in}} \times 100 \%
\end{equation}

For practical reasons, we assume that the system state, input, output and uncertainty are restricted to be in the compact sets $\m X$, $\m U$, $\m Y$ and $\m W$ respectively.

\subsection{Control problem formulation}

In a post combustion carbon dioxide capture plant, the primary objective of the absorption column is to reduce the amount of carbon dioxide in the flue gas emanating from the power plant. This can be achieved by controlling the efficiency $y$ of the absorption column. The absorption efficiency can be controlled by manipulating the lean solvent flow rate $F_L$ and the concentration of CO$_2$ in the lean solvent entering the top of the column. In this work, the concentration of CO$_2$ in the lean solvent is kept constant since the desorption column is not considered. Therefore, only the lean solvent flow rate is used as the manipulated variable. Under these conditions, a very high CO$_2$ absorption efficiency which translates to high removal of CO$_2$ from the inlet flue gas may be achieved by using a high lean solvent flow rate. However, using a high amount of solvent to reduce the amount of CO$_2$ in the flue gas can have negative effects on the operation of the absorption column and the PCC plant as a whole. First, a high amount of solvent may cause flooding in the column. Column flooding is usually followed by a dramatic increase in column pressure and prevent the flue gas from flowing out of the column. This may result in inefficient operation of the column and/or equipment damage. Second, a high solvent flow rate may often than not lead to overcirculation of the solvent in the PCC plant. This usually results in the solvent leaving the absorption column (rich solvent) having a low CO$_2$ concentration. This makes it difficult to operate the desorption column efficiently. Therefore the desire is usually to keep the absorption efficiency $y$ within a target zone $\m Y_t$ which ensures a balance between high absorption efficiency, column flooding and overcirculation.

The primary control objective of this work is therefore to design a feedback controller which drives the system state to a predetermined target zone $\m Y_t$ if the initial absorption efficiency of the system is outside the target zone and subsequently maintain the state of the system within the target zone $\m Y_t$ thereafter. A secondary objective is to minimize the average economic cost $\ell_e$ over an infinite horizon $T$ which is given by

\begin{equation} \label{eqn:average_cost}
    \limsup_{T \rightarrow \infty} \frac{1}{T} \sum_{t=0}^{T-1} \ell_e(y(t))
\end{equation}
where 
\begin{equation}\label{eqn:economic_objective}
    \ell_e(y) = -y
\end{equation}
denotes the economic objective to be minimized. To achieve the above control objectives, we resort to EMPC with zone tracking \cite{liu2018,DECARDI_NELSON2022} and implicitly take into account the presence of process disturbance $w$ in the design of the EMPC. 


\section{Economic model predictive control with zone tracking}

In this section, we present the economic model predictive control with zone tracking (ZEMPC) algorithm. Specifically, two variations of the ZEMPC algorithm are presented. The first control algorithm denoted as nominal ZEMPC (NZEMPC) is the economic model predictive control with target zone tracking. In this formulation, the disturbances are not considered in the design. This serves as a basis to compare the second variation of ZEMPC. In the second control algorithm denoted as robust ZEMPC (RZEMPC), the original target zone $\m Y_t$ is modified to implicitly consider the effects of the uncertainty in the process system.

We begin this section by presenting the NZEMPC formulation. Thereafter, an algorithm to modify the original target zone $\m Y_t$ is presented. Finally, we present the RZEMPC formulation. 

\subsection{Economic MPC with target zone tracking}
Given information about the current state $x(t_k)$ at sampling time $t_k$, the  ZEMPC uses the nominal model of System~(\ref{eqn:discretized_model}):
\begin{equation} \label{eqn:nominal_system}
    \dot{\tilde{x}} (t) = f(\tilde x (t),v(t))
\end{equation}  
with the initial condition $\tilde x (t_k) = x(t_k)$ to find a control sequence $ \textbf{v}=\{ v(t_k), \hdots, v(t_{k}+N\Delta)\}$ and the associated state sequence $\mathbf{\tilde{x}}=\{ \tilde x (t_k), \hdots, \tilde x (t_k + \Delta N) \}$ over the entire prediction horizon $N$ at a sampling time $\Delta$ to minimize the cost functional:
\begin{equation} \label{eqn:cost_function}
    V_N(x(t_k),\textbf{v}) = \int_{t_k}^{t_k + N\Delta} \ell(y(t)) dt 
\end{equation}
In Equations~(\ref{eqn:nominal_system}) and (\ref{eqn:cost_function}), $\tilde{x}(t) \in \m{X} \subseteq \m{R}^{50}$ and $v(t) \in \m{U} \subseteq \m{R}$ are the nominal state vector and computed control input vector respectively. The stage cost $\ell(\cdot)$ is defined as follows: 
\begin{equation} \label{eqn:stage_cost}
    \ell(y) = \ell_e(y) + \ell_z(y)
\end{equation}
where $\ell_e(\cdot)$ is the economic stage cost as introduced in (\ref{eqn:economic_objective}) and $\ell_z(\cdot)$ is a zone tracking penalty term which is defined as below:
\begin{subequations} \label{eqn:zone_penalty_opt}
    \begin{align} 
        \ell_z(y) = \min_{y_z} & ~~~ c_1(\| y - y_z \|_2^2) \label{eqn:zone_penalty_opt_a}\\
        s.t. & ~~~ y_z \in \m{Y}_t \label{eqn:zone_penalty_opt_b}
    \end{align}
\end{subequations}
with $c_1 \in \m{R}_{+}$ being a non-negative weight on the zone tracking term and $y_z$ is a slack variable. The zone tracking stage cost reflects the distance of the system states from the target zone and is positive definite. The weights $c_1$ must be appropriately selected such that the zone tracking cost is given a higher priority than the economic objective. 

At each sampling time, the following dynamic optimization problem $\mathcal{P}_N(x(t_k))$ is solved:
\begin{subequations} \label{eqn:zone_opt}
    \begin{align} 
        \min_{\bf{v},\bf{y_z}} & ~~~ \int_{t_k}^{t_k + N\Delta} -y(t) + c_1(\| y(t) - y_z(t) \|_2^2) dt \label{eqn:zone_opt_a}\\
        s.t. & ~~~ \dot{\tilde{x}}(t) = f(\tilde{x}(t),v(t)) \label{eqn:zone_opt_b} \\
        & ~~ y(t) = h(\tilde{x}(t)) \label{eqn:zone_opt_c}\\
        & ~~~ \tilde{x}(t_{k}) = x(t_{k}) \label{eqn:zone_opt_d}\\
        & ~~~  \tilde{x}(t) \in \m{X} \label{eqn:zone_opt_e}\\
        & ~~~ v(t) \in \m{U}  \label{eqn:zone_opt_f}\\
        & ~~~ y(t) \in \m{Y} \label{eqn:zone_opt_g} \\
        & ~~~ y_z(t) \in \m{Y}_t \label{eqn:zone_opt_h}
    \end{align}
\end{subequations}
In the Optimization problem~(\ref{eqn:zone_opt}) above, Equation~(\ref{eqn:zone_opt_b}) is the model constraint, Equation~(\ref{eqn:zone_opt_c}) represents the output relationship,  Equation~(\ref{eqn:zone_opt_d}) is the initial state constraint, Equation~(\ref{eqn:zone_opt_e}) -- (\ref{eqn:zone_opt_g}) are the constraints on the state, input and output respectively, and Equation~(\ref{eqn:zone_opt_h}) is the zone constraint. As a result of the cost function employed, the Optimization problem~(\ref{eqn:zone_opt}) is a multi-objective optimization problem which seeks to minimize the deviation of the absorption efficiency from the target zone $\m{Y}_t$ while at the same time maximizing the the efficiency within the target zone.

The solution of $\mathcal{P}_N(x(t_k))$ denoted $\mathbf{v^*}$ gives  an optimal value of the cost $V_N^0 (x(t_k))$ and at the same time $u(t_k)=v^*(t_{k})$ is applied to the actual system (\ref{eqn:discretized_model}).

\subsection{Modification of the target zone}

While the NZEMPC described in the earlier section ensures that the zone tracking objective is achieved for the absorption column without any uncertainty, the zone tracking objective may not be achieved in general for systems with uncertainty.  This may be due to two reasons. First, the target zone may not necessarily be forward invariant for the closed-loop system. This means that it is possible for the uncertainty to drive the system's states to a region outside the forward invariant set but within the target zone. Once the state is outside the forward invariant set, the target zone cannot be tracked anymore and will ultimately result in system instability. Second, the NZEMPC algorithm may cause the system output to operate very close to the boundary of the target zone due to the secondary economic objective employed; that is maximization of the absorption efficiency. While this may not be an issue in the nominal case, the presence of the uncertainty may drive the system states outside the target zone making it difficult to track the zone.

Therefore to achieve the zone tracking objective in the presence of uncertainty, we modify the target zone used in the formulation of the EMPC with zone tracking algorithm. It is worth mentioning that modification of the target zone is not trivial. For example, merely selecting the center of the zone as the target zone to track may not be best choice. This will be demonstrated in the simulation section. The idea is to find a robust control invariant set within the target zone which ensures that the target zone can still be tracked in the presence of the uncertainty. A robust control invariant set $R$ is a set of initial states in the target zone for which there exists a control action such that the trajectory of the system stays in $R$ for all future times irrespective of the disturbances. Ideally, the largest robust control invariant set within the target zone is desired \cite{DECARDI_NELSON2022}. However, finding the largest robust control invariant sets for large scale systems is very difficult and the method proposed by Decardi-Nelson and coworkers \cite{DECARDI_NELSON2022} cannot be applied to the absorption column due to the number of states involved. We therefore resort to using simpler ellipsoidal control invariant set techniques. Since the ellipsoidal control invariant set does not consider the presence of uncertainty in the control system, we use a back-off approach not only to account for the presence of disturbances but to also avoid operating close to the boundary of the target zone $\m Y_t$.

Before we present the zone modification algorithm, let us first define the steady-state (SS) optimization problem with respect to the target zone as

\begin{subequations} \label{eqn:ss_opt}
    \begin{align}
        (\ell_e^*,x_s,u_s) &= \arg \min ~ \ell_e(y) \label{eqn:ss_opt_cost}\\
        s.t. ~~~ & 0 = f(x,u) \label{eqn:ss_opt_model}\\ 
        & y = h(x) \label{eqn:ss_opt_output}\\ 
        & x \in \m{X} \label{eqn:ss_input_con}\\
        & u \in \m{U} \label{eqn:ss_opt_state_con} \\
        & y \in \m Y_t \label{eqn:ss_opt_output_con}
    \end{align}
\end{subequations}
In Optimization problem~(\ref{eqn:ss_opt}), Equation~(\ref{eqn:ss_opt_model}) is the system model defined in System (\ref{eqn:discretized_model}) without any uncertainty, and Equation~(\ref{eqn:ss_opt_output}) -- (\ref{eqn:ss_opt_output_con}) are the same as defined previously. The Optimization problem~(\ref{eqn:ss_opt}) returns the optimal economic cost $\ell_e^*$ within the target zone $\m Y_t$ as well as the steady-state state $x_s$ and input $u_s$. The $\ell_e^*$ serves as a lower bound on the economic cost that can be achieved in the target zone $\m Y_t$. In the proposed zone modification algorithm, this value is relaxed by multiplying it with the relaxation rate $r$ to obtain the relaxed optimal economic cost $\ell_e^r$. Here, relaxation of the optimal steady-state economic cost within the target zone means increasing the value of the economic cost above $\ell_e^*$ such that 
\begin{equation}
    \ell_e^* < \ell_e^r
\end{equation}
Relaxing the best economic cost within the target zone sacrifices some economic performance to implicitly account for the effects of the uncertainty in the control system. To achieve the relaxation, the following cost-relaxed steady-state (CRSS) optimization problem is solved

\begin{subequations} \label{eqn:relaxed_ss_opt}
    \begin{align}
        (x_s,u_s) &= \arg \min ~ 0 \label{eqn:relaxed_ss_opt_cost}\\
        s.t. ~~~ & 0 = f(x,u) \label{eqn:relaxed_ss_opt_model}\\ 
        & x \in \m{X} \label{eqn:relaxed_ss_opt_state_con}\\
        & u \in \m{U} \label{eqn:relaxed_ss_opt_input_con}\\
        & y \in \m Y_t \label{eqn:relaxed_ss_opt_output_con}\\
        & \ell_e(y) = \ell_e^r \label{eqn:relaxed_ss_opt_cost_con}
    \end{align}
\end{subequations}
In Optimization problem~(\ref{eqn:relaxed_ss_opt}) above, Equation~(\ref{eqn:relaxed_ss_opt_cost_con}) is the economic cost relaxation constraint which needs to be achieved. It can be seen that Optimization problem~(\ref{eqn:relaxed_ss_opt}) is a feasibility problem since it does not seek to minimize any cost function. The CRSS problem returns the steady-state operating point ($x_s$,$u_s$) at the relaxed economic cost function value. This is used in the subsequent parts of the zone modification algorithm to obtain the ellipsoidal control invariant set.

To compute the ellipsoidal control invariant set, the nonlinear system is first linearized about the steady-state operating point ($x_s$,$u_s$) from Optimization problem~(\ref{eqn:relaxed_ss_opt}) to obtain a linear system of differential equations of the form
\begin{equation}\label{eqn:linear_model}
    \dot{\bar{x}}(t) = A\bar{x}(t) + B\bar{u}(t)
\end{equation}
where $A = \frac{\partial f(x,u)}{\partial x} |_{x_s,u_s} $ and $B = \frac{\partial f(x,u)}{\partial u} |_{x_s,u_s}$ are matrices of appropriate dimensions, and $\bar{x}$ and $\bar{u}$ denote the system states and input in deviation form i.e. $\bar{x} = x - x_s$ and $\bar{u} = u - u_s$. Following the linearization, a semi-definite program (SDP) is formulated according to the work by Polyak and Shcherbakov \cite{POLYAK2009}. The SDP to be solved is presented in Optimization problem~(\ref{eqn:ellipsoid_opt}).
\begin{subequations} \label{eqn:ellipsoid_opt}
    \begin{align} 
        \max_{P,Y} & ~~~ \text{trace}(P) \\
        s.t. & ~~~ AP + PA^T + BY + Y^TB^T \prec 0 \\
        & ~~~ \begin{bmatrix}
            P & Y^T \\
            Y & \bar{u}_{\text{max}}^2 I
        \end{bmatrix} \succeq 0  
    \end{align}
\end{subequations}
where $P$ and $Y$ are matrices of appropriate dimension and $\bar{u}_{\text{max}}$ is the bound on the input i.e. $\| \bar{u} \| \leq \bar{u}_{\text{max}}$. By maximizing the trace of $P$, the maximal ellipsoidal control invariant set under the constrained input can be obtained from the optimal solution of (\ref{eqn:ellipsoid_opt}) as 
\begin{equation}
    \m X_m = \{ \bar{x} \in \m{R}^{50} : \bar{x}^T P^{-1} \bar{x} \leq 1 \}, \quad \quad P \succ 0
\end{equation}
It is worth mentioning that computing an ellipsoidal control invariant set for the case where System \eqref{eqn:linear_model} is stable (i.e. real part of the eigen values of $A$ are negative) is trivial. This is because the stability region spans the entire state space and the input is not necessary for stabilization. Thus, Optimization problem~\eqref{eqn:ellipsoid_opt} will not return a solution since the maximal ellipsoidal control invariant set is unbounded. In such a situation, the Lyapunov equation $AP + PA^T + Q = 0$ is solved with $Q$ being an identity matrix of appropriate dimension. In this case the ellipsoidal control invariant set is obtained as $\m X_m = \{ \bar{x} \in \m{R}^{50} : \bar{x}^T P \bar{x} \leq \alpha \}$ where $\alpha \geq 0$ is a scalar parameter which determines the size of the invariant set.

\begin{remark}
    A linear system obtained by linearizing a nonlinear system might only be accurate within a very small region of the linearization point (origin). To avoid obtaining an ellipsoidal control invariant set that spans areas in the state space for which the linear model is inaccurate, the bound on the input $u_{\text{max}}$ should be reduced. The magnitude of the reduction ultimately depends on the properties of the system.
\end{remark}

\begin{figure}[tbp]
  \begin{center}
    \includegraphics[width=0.5\textwidth]{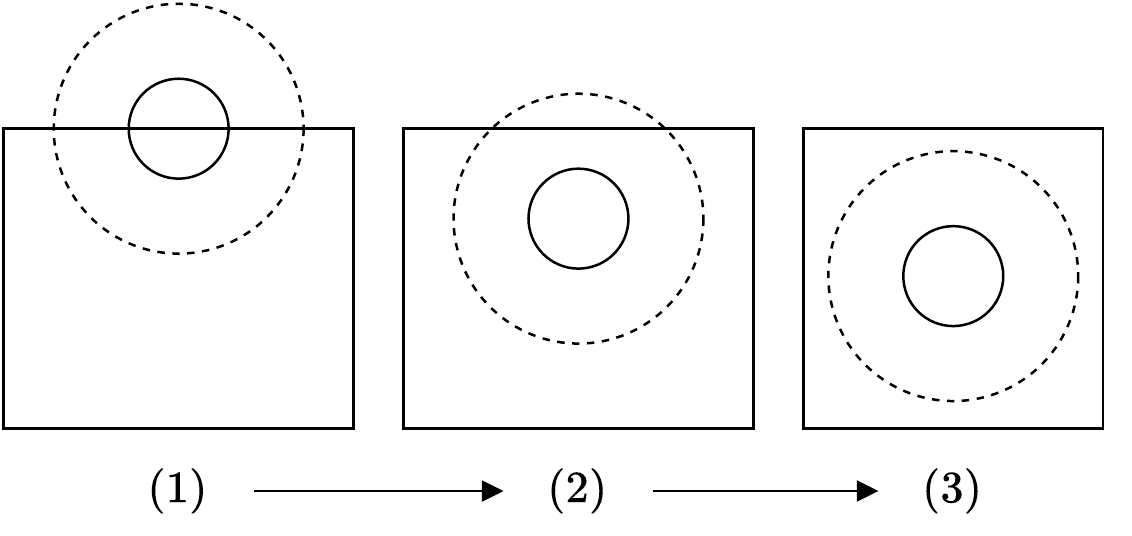}    
    \caption{An illustration of three iterations of the zone modification algorithm in a fictitious two dimensional space. The rectangle represents the original target zone, the circle with solid line represent the ellipsoidal invariant set and the circle with dashed lines represent the $\varepsilon \m B$ enlargement of the ellipsoidal invariant set. The algorithm terminates in the third step when the $\varepsilon \m B$-enlarged ellipsoidal invariant set does not intersect with the set outside the target zone.} 
    \label{fig:zone_mod_algorithm}
  \end{center}
\end{figure}

Once $\m X_m$ is obtained, it is projected into the output space using the output equation to obtain the modified target zone $\m Y_m$ such that
\begin{equation}\label{eqn:output_set}
    \m Y_m = h(\m X_m)
\end{equation}
Since the output equation in Equation~(\ref{eqn:output}) depends on only the concentration of CO$_2$ in the gas exiting the top of the absorption column (inlet CO$_2$ concentration in the gas phase is fixed), the minimum and the maximum state in $\m X_m$ can be used to obtain $\m Y_m$. For more general cases, a finite sample of states in $\m X_m$ may be required. The modified output zone $\m Y_m$ is then enlarged by an $\varepsilon$-ball. This is used as a stopping criterion by checking if the enlarged modified output target zone ($\m Y_m + \varepsilon \m B$) does not intersect with any part of the output space outside the output target zone i.e.
\begin{equation}\label{eqn:stop_criterion}
    \m Y_m + \varepsilon \m B \cap \m Y_t \backslash \m Y = \emptyset
\end{equation}
The procedure is run in while loop until the stopping criterion in Equation~(\ref{eqn:stop_criterion}) is met. The entire zone modification algorithm is summarized in Algorithm \ref{alg:ezone}. A visual depiction of the algorithm is shown in Figure~\ref{fig:zone_mod_algorithm}.
The algorithm takes as inputs the nominal system model $f$, the output equation $h$, the state $\m X$ and input $\m U$ constraints, the optimal economic cost within the target zone $\ell_e^*$ as well as the parameters $\varepsilon$, the cost relaxation rate $r$ and the maximum number of iterations $N_{\text{max}}$. $N_{\text{max}}$ is introduced in the algorithm to ensure that the while loop does not run indefinitely if a poor choice of the parameters are selected. The algorithm terminates with an empty set if the parameters are poorly chosen.

\begin{algorithm}[tbp] \label{alg:ezone}
    \caption{Modification of target zone}
    \KwIn{ $f$, $h$, $\ell_e$, $\m X$, $\m{U}$, $\m Y_t$, $u_{\text{max}}$, $\varepsilon$, $r$, $N_{\text{max}}$, $\ell_e^*$}
    \KwOut{ $\m{X}_m$, $\m Y_m$ }

    $\m Y_0 \leftarrow \m Y_t$ \\
    $\m X_0 \leftarrow \emptyset $ \\
    $\ell_e^r \leftarrow \ell_e^*$ \\
    $i \leftarrow 1$ \\
    
    \While{$\m Y_{i-1} + \varepsilon \m B \cap \m Y_t \backslash \m Y$}{
        $\ell_e^r \leftarrow (1 + r) \ell_e^r $ \\
        Solve the optimization problem in (\ref{eqn:relaxed_ss_opt}) to obtain ($x_s,u_s$)\\
        Linearize the System (\ref{eqn:discretized_model}) at ($x_s,u_s$) to obtain $A$ and $B$ \\
        \eIf{$A$ is stable}{
        Solve the Lyapunov equation to obtain $P$
        }{
        Solve \eqref{eqn:ellipsoid_opt} to obtain $P$
        }
        
        Find the ellipsoidal control invariant set $\m X_i$ using $P$ \\
        $\m Y_i \leftarrow h(\m X_i)$ \\
        \If{$ i = N_{\text{max}}$}{
            $\m X_i \leftarrow \emptyset$ \\
            $\m Y_i \leftarrow \emptyset$ \\
            \textbf{break}
        }

        $i \leftarrow i + 1$
    } 
    
    $\m{X}_m \leftarrow \m X_i$ \\ 
    $\m{Y}_m \leftarrow \m Y_i$ \\ 
    
    \textbf{return} $\m{X}_m$, $\m Y_m$

\end{algorithm}

While the modified output zone $\m Y_m$ is in a form which can be used in the NZEMPC optimization (\ref{eqn:zone_opt}), this may not the best approach. This is because there is no guarantee that the set $\m Y_m$ obtained from the projection of the ellipsoidal control invariant set $\m X_m$ is also control invariant. Therefore, the ellipsoidal control invariant set $\m X_m$ needs to be used the MPC algorithm. However, replacing the zone slack constraint (\ref{eqn:zone_opt_h}) in Optimization problem~(\ref{eqn:zone_opt}) will lead to having the number of slack variables equal to the dimension of the state of the system. Since in most control systems the dimension of the output is smaller than or equal to the dimension of the system states, the increased number of slack variables $y_z$ will eventually increase the size of the optimization problem to be solved online. Thus to mitigate this problem, the zone constraint is modified to Equation~(\ref{eqn:modified_zone_opt_h}) and an additional positivity constraint (\ref{eqn:modified_zone_opt_i}) is added. This results in only one slack variable with is independent of the input or the output dimension. Full details of the robust economic Model Predictive Control with zone tracking (RZEMPC) algorithm is presented in Optimization problem~(\ref{eqn:modified_zone_opt}). 
\begin{subequations} \label{eqn:modified_zone_opt}
    \begin{align} 
        \min_{\bf{v},\bf{y_z}} & ~~~ \int_{t_k}^{t_k + N\Delta} -y(t) + c_1(\| y_z(t) \|_2^2) dt \label{eqn:modified_zone_opt_a}\\
        s.t. & ~~~ \dot{\tilde{x}}(t) = f(\tilde{x}(t),v(t)) \label{eqn:modified_zone_opt_b} \\
        & ~~ y(t) = h(\tilde{x}(t)) \label{eqn:modified_zone_opt_c}\\
        & ~~~ \tilde{x}(t_{k}) = x(t_{k}) \label{eqn:modified_zone_opt_d}\\
        & ~~~  \tilde{x}(t) \in \m{X} \label{eqn:modified_zone_opt_e}\\
        & ~~~ v(t) \in \m{U}  \label{eqn:modified_zone_opt_f}\\
        & ~~~ y(t) \in \m{Y} \label{eqn:modified_zone_opt_g} \\
        & ~~~ (\tilde{x}(t)-x_s)^T P (\tilde{x}(t) - x_s) \leq 1 + y_z(t) \label{eqn:modified_zone_opt_h} \\
        & ~~~  y_z(t) \geq 0 \label{eqn:modified_zone_opt_i}
    \end{align}
\end{subequations}
The constraints in Optimization problem~(\ref{eqn:modified_zone_opt}) are the same as previously defined.

\section{Simulation results}
In this section, we carry out different set of simulations to demonstrate the effectiveness and applicability of the economic MPC with modified zone tracking to the absorption column of the PCC plant. We also compare the results of the control algorithm with the modified zone to that of the controller with the original target zone. Specifically, we present the performance of the controllers under additive state uncertainties which represent various kinds of uncertainties in the process model. We also compare the performance of the controllers under time-varying inlet flue gas flow rate.

 We begin this section with the simulation settings and model parameters used in this work. Then, we investigate the effects of additive state uncertainty that may be present in the process. Finally, we analyze and compare the performance of the controllers when the inlet flue gas flow rate vary.

\subsection{Simulation settings}

The plant configuration for the process model described in Section \ref{subsec:process_description} is determined according to Decardi-Nelson et al. \cite{DECARDI_NELSON2018} and is presented in Table~\ref{tb:unit_configuration}.
\begin{table}[tbp]
  \begin{center}
  \caption{CO$_2$ gas absorption column configuration}\label{tb:unit_configuration}
  \vspace{2mm}
    \begin{tabular}{lc}
        \hline
    Property & Value \\  
        \hline
    Column internal diameter $D_c$ (m) & $0.43$ \\ 
    Packing height (m) & $6.1$ \\ 
    Packing type & IMTP \#40 \\
    Nominal packing size (m) & $0.038$ \\
    Specific packing area (m$^2$) & 143.9 \\
        \hline
    \end{tabular}
  \end{center}
\end{table}

\begin{table}[tp!]
  \begin{center}
  \caption{Nominal flue gas condition}\label{tb:flue_gas_condition}
  \vspace{2mm}
    \begin{tabular}{lc}
        \hline
    Property & Value \\  
        \hline
    Temperature (K) & $319.70$ \\ 
    Volumetric Flow rate $F_G$ (m$^3$/s) & $0.0832$ \\ 
    CO$_2$ mole fraction & $0.1500$ \\
    N$_2$ mole fraction & $0.8000$ \\
    MEA mole fraction & $0.0000$ \\
    H$_2$O mole fraction & $0.0500$ \\
        \hline
    \end{tabular}
  \end{center}
\end{table}

\begin{table}[tp!]
  \begin{center}
  \caption{Nominal inlet amine solvent condition}\label{tb:solvent_condition}
  \vspace{2mm}
    \begin{tabular}{lc}
        \hline
    Property & Value \\  
        \hline
    Temperature (K) & $314.0$ \\ 
    CO$_2$ mole fraction & $0.0266$ \\
    N$_2$ mole fraction & $0.0000$ \\
    MEA mole fraction & $0.1104$ \\
    H$_2$O mole fraction & $0.8630$ \\
        \hline
    \end{tabular}
  \end{center}
\end{table}

The properties of the inlet flue gas entering the column from the bottom and the inlet solvent entering the column from the top are also shown in Tables \ref{tb:flue_gas_condition} and \ref{tb:solvent_condition} respectively. In this work, we assume that the properties of the solvent entering the absorption column from the desorption unit is fixed with the exception of the flow rate which is manipulated. The flue gas flow rate to the column may vary but this is unknown to the controller.

In this work, we assume that all the states are measured and available to the controller at any sampling time $t_{k \geq 0}$ with a sampling interval $\Delta$ set at $10$ minutes. This is a reasonable assumption since it has been shown that only the temperature measurements, which can be easily obtained, can be used to reconstruct the full states of the absorption column \cite{YIN2019}. Unless otherwise stated, the prediction and control horizon $N$ of the controllers is set at 10. The parameters $\varepsilon$, $r$, $\alpha$ and $N_{\text{max}}$ in Algorithm \ref{alg:ezone} are fixed at 0.009, -0.005, 5, and 10 respectively. The value of $u_{\text{max}}$ was fixed at 20 \% of the available input energy for control and the identity matrix was used as the value of $Q$.  The output zone to be tracked was chosen to be between 0.85 and 0.90 i.e. $\m Y_z = [0.85 ~ 0.90]$ with a target zone tracking weight $c_1 = 10000$. This ensures that a high absorption efficiency is not pursued by the controller due to the economic objective which can cause operational issues such as flooding of the column and solvent over-circulation at high liquid flow rates. This also prevents the controller from allowing large quantities of CO$_2$ to be released into the atmosphere resulting in higher CO$_2$ taxes. In the implementation of the control algorithms, the system states were scaled such that 
\begin{equation}
    \hat{x}(t_k) = x(t)/x_{\text{scale}}, \quad \hat{u}(t_k) = u(t)/u_{\text{scale}}
 \end{equation} 
 where $\hat{x}$ and $\hat{u}$ are the scaled states and inputs respectively, and $x_{\text{scale}}$ and $u_{\text{scale}}$ are the steady-state values corresponding to the center of the zone i.e. an absorption efficiency of 0.875.
The additive state uncertainty $w$ in \eqref{eqn:discretized_model} was assumed to be uniformly distributed in $[-0.00001 \times \textbf{1} ~ 0.00001 \times \textbf{1}]$ where $\textbf{1} = x_{\text{scale}}/x_{\text{scale}}$ is a vector of ones having the same dimension as the state. 

\subsection{Results and discussion}
In this section, we present the results for the two control algorithms namely EMPC with target zone tracking (NZEMPC) and EMPC with modified zone tracking (RZEMPC). We also consider the case where the modified zone is at the center of the target zone. This case was added to illustrate the notion that arbitrarily tracking a zone within the center of the target zone, though easy, may not be the best strategy to ensure finite-time zone tracking with good performance. Performance or cost as defined in this section refers to Equation~\eqref{eqn:stage_cost} which represents both the economic performance and the ability of the controller to ensure that the output is within target zone at all times.

\subsubsection{Additive state uncertainty}
The average performance of the controllers for the case of additive state disturbances are shown in Table~\ref{tb:result_comparison}. As can be seen from Table~\ref{tb:result_comparison}, our proposed zone EMPC control algorithm with modified target zone outperforms that of the controller that tracks the original target zone. The EMPC with modified target zone at the center of the zone performs better than tracking the original target zone but does not perform better than that with the modified target zone. This implies that the tuning parameters in Algorithm \ref{alg:ezone} needs to be carefully selected to ensure that there's both performance and zone tracking objectives are achieved. One way to do this is to consider that uncertainty information when selecting $\varepsilon$ as this ensures a reasonable back-off from the boundary of the target zone.
\begin{table}[tp!]
  \begin{center}
  \caption{Comparison of the nominal EMPC with target zone tracking and the two EMPCs with modified zone tracking (smaller is better) }\label{tb:result_comparison}
  \vspace{2mm}
    \begin{tabular}{lc}
        \hline
    Zone & Average cost \\  
        \hline
    Target zone & $ -0.72785$ \\ 
    Modified zone  & $-0.88639$ \\ 
    Center of zone & $-0.86811$ \\
        \hline
    \end{tabular}
  \end{center}
\end{table}

\begin{figure}[tbp]
  \begin{center}
    \includegraphics[width=0.6\textwidth]{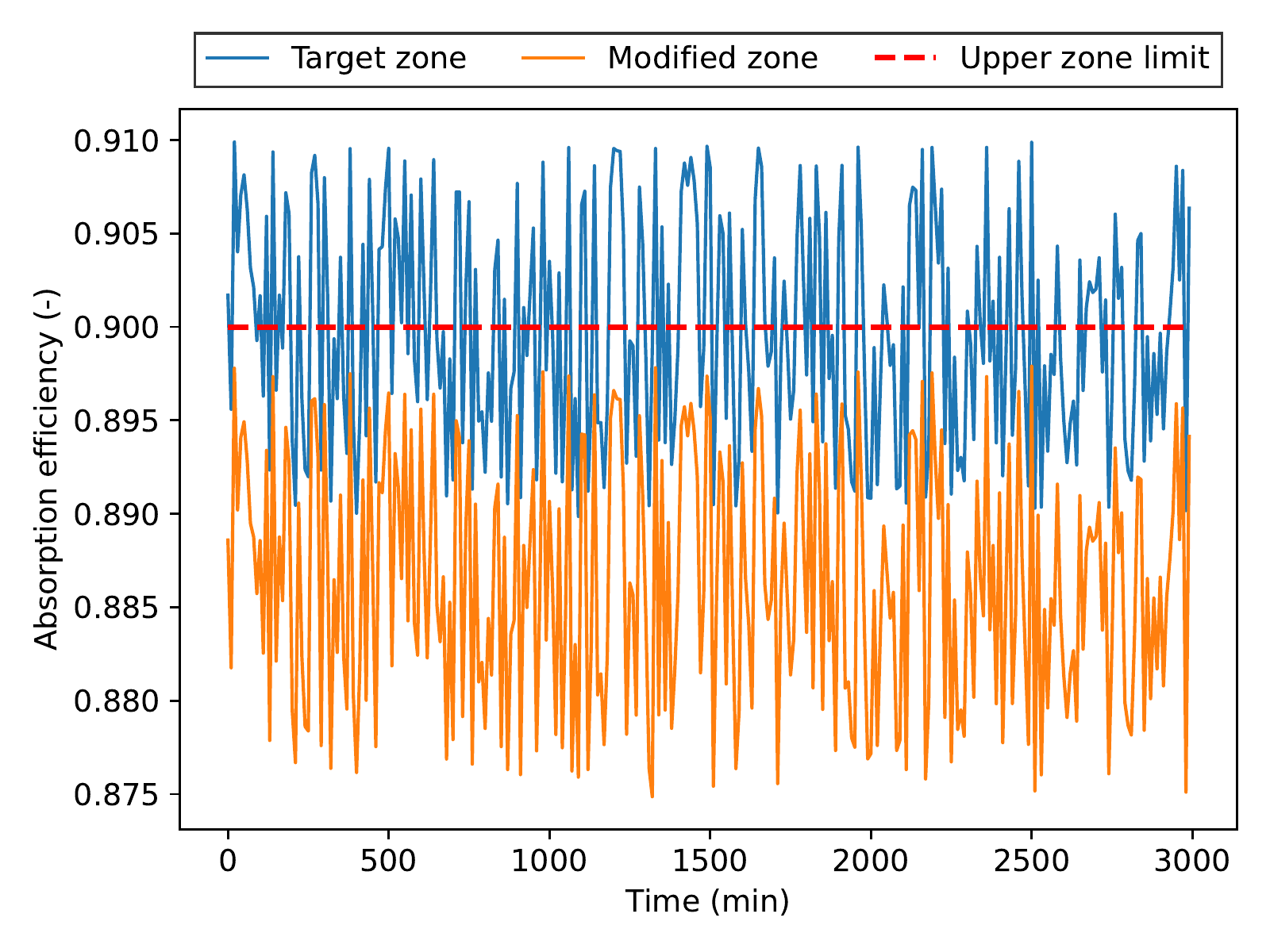}    
    \caption{Trajectories of the absorption efficiency for the absorption column under the operation of the zone EMPC control algorithm tracking the original target zone (blue) and modified zone (orange)} 
    \label{fig:additive_state_uncertainty}
  \end{center}
\end{figure}

\begin{figure}[tbp]
  \begin{center}
    \includegraphics[width=0.6\textwidth]{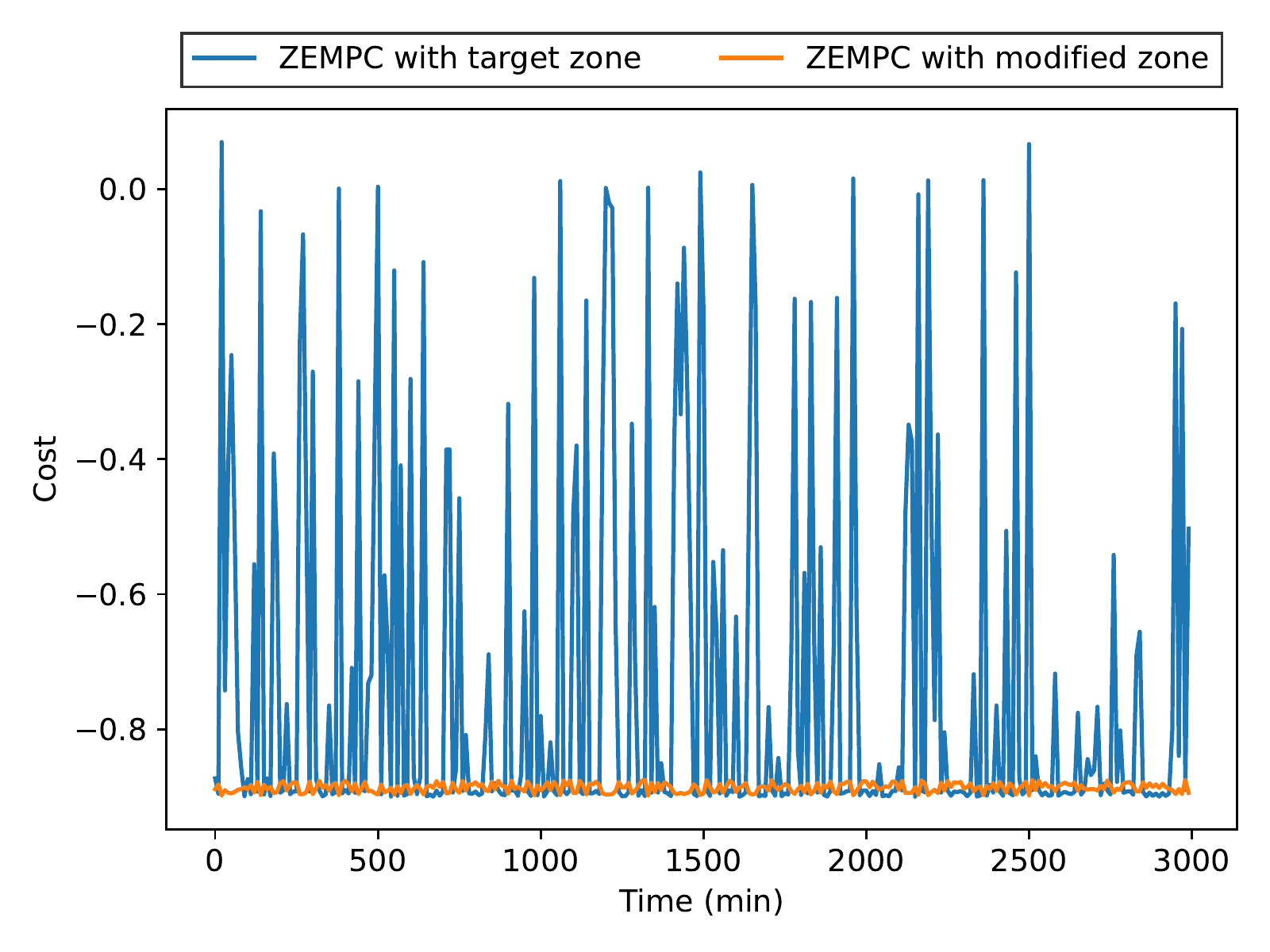}    
    \caption{Trajectories of the stage cost for the absorption column under the operation of the zone EMPC control algorithm tracking the original target zone (blue) and modified zone (orange)} 
    \label{fig:cost}
  \end{center}
\end{figure}

To understand why this happens, Figures~\ref{fig:additive_state_uncertainty} and \ref{fig:cost} have been provided. As mentioned earlier, the presence of the secondary economic objective can cause the system output to operate close to the boundary of the target zone. This is the case for the operation of the absorption column since the economic objective is to maximize the absorption efficiency. The presence of the uncertainty causes the system output to move out of the target leading to high cost and inability to track the target zone. By modifying the target zone, our proposed controller ensures that there is room available for the system to operate within the target zone even in the presence of the uncertainty. It is worth mentioning that, by modifying the target zone, some performance is sacrificed in favour of achieving the zone tracking objective. This can be seen in Figure~\ref{fig:additive_state_uncertainty} where the controller tracking the target zone operates at a higher absorption efficiency compared to the one that tracks the modified target zone. Because of the target zone violation, it can be seen in Figure~\ref{fig:cost} that the cost trajectory of the NZEMPC is erratic compared to that of the RZEMPC.

\subsubsection{Time-varying flue gas flow rate} 
The operation of a typical power plant is usually periodic every day and seasonally. This is because of the variation in electricity demand. Electricity demand is usually low in the early morning and very late at night where consumer activity is low. It gradually rises to a peak around noon and stays there for sometime before finally reducing again at night. Furthermore, it has been suggested that renewable energy sources be integrated into the energy generation mix with the renewable energy sources being the main power generation sources and the fossil fuel power plants as backups. However, some renewable energy sources may not be very reliable. For example, the ability of a solar panel to generate electricity depends on the availability of sunlight which may not always be available. This integrated energy mix will therefore further cause more erratic operation of the fossil fuel power plant. A consequence of this changes in the demand and subsequently the output of the power plant is that the flue gas emanating from the power plant to the absorption column will vary frequently. This time-varying behaviour can have significant effects on the performance of the PCC plant attached to the fossil fueled power plant. Therefore, flexible operation of the PCC plant attached to the power plant is inevitable. This has been the subject of several studies in the control of PCC plant attached to a load-following power plant \cite{BANKOLE2018,PATRON2020,DECARDI_NELSON2018,RUA2020}

\begin{figure}[tp!]
  \begin{center}
    \includegraphics[width=0.6\textwidth]{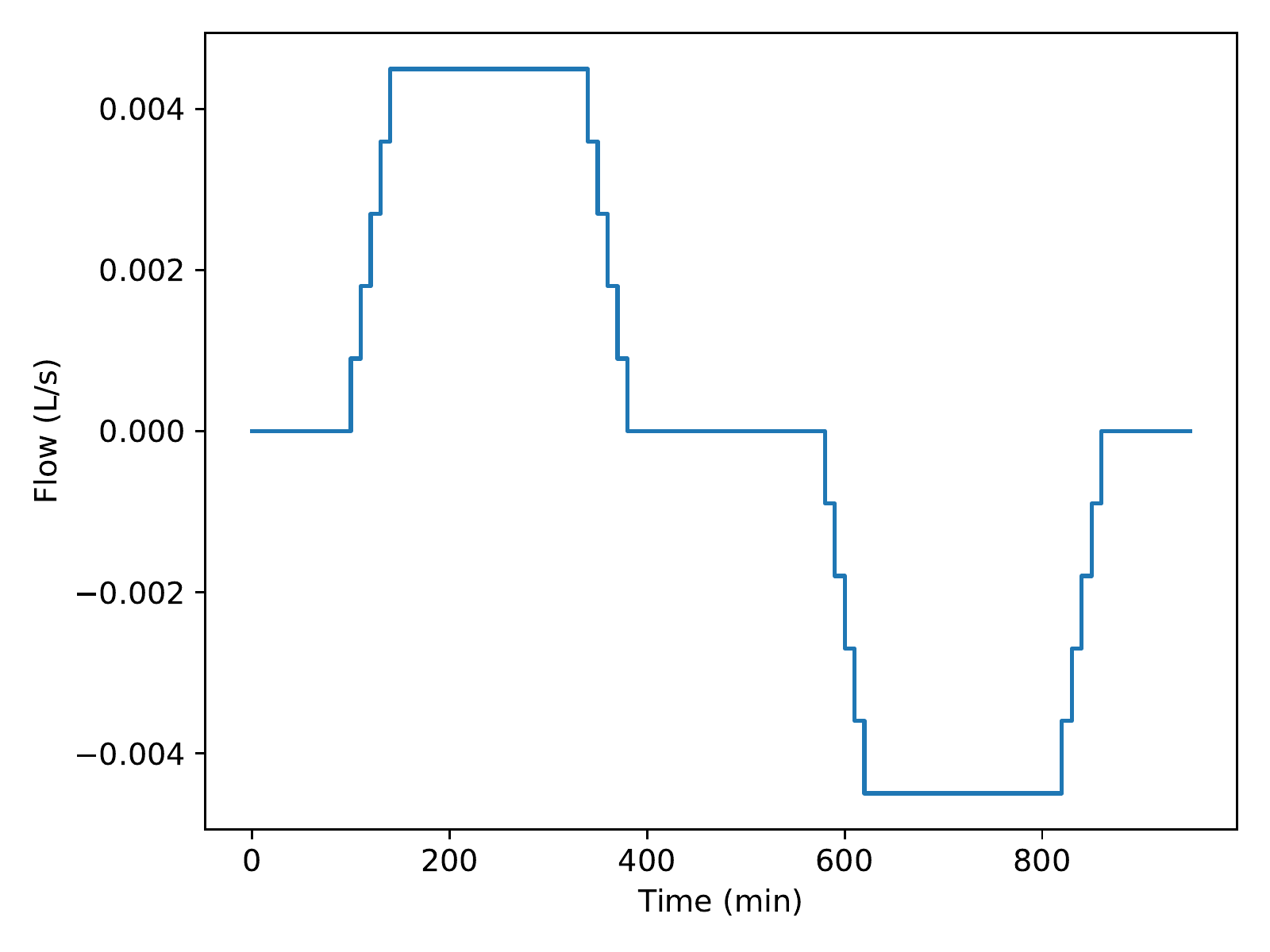}    
    \caption{Trajectories of the disturbance of the flue gas flow rate signifying ramping up and ramping down operations} 
    \label{fig:tvf_flow}
  \end{center}
\end{figure}

\begin{figure}[tp!]
  \begin{center}
    \includegraphics[width=0.6\textwidth]{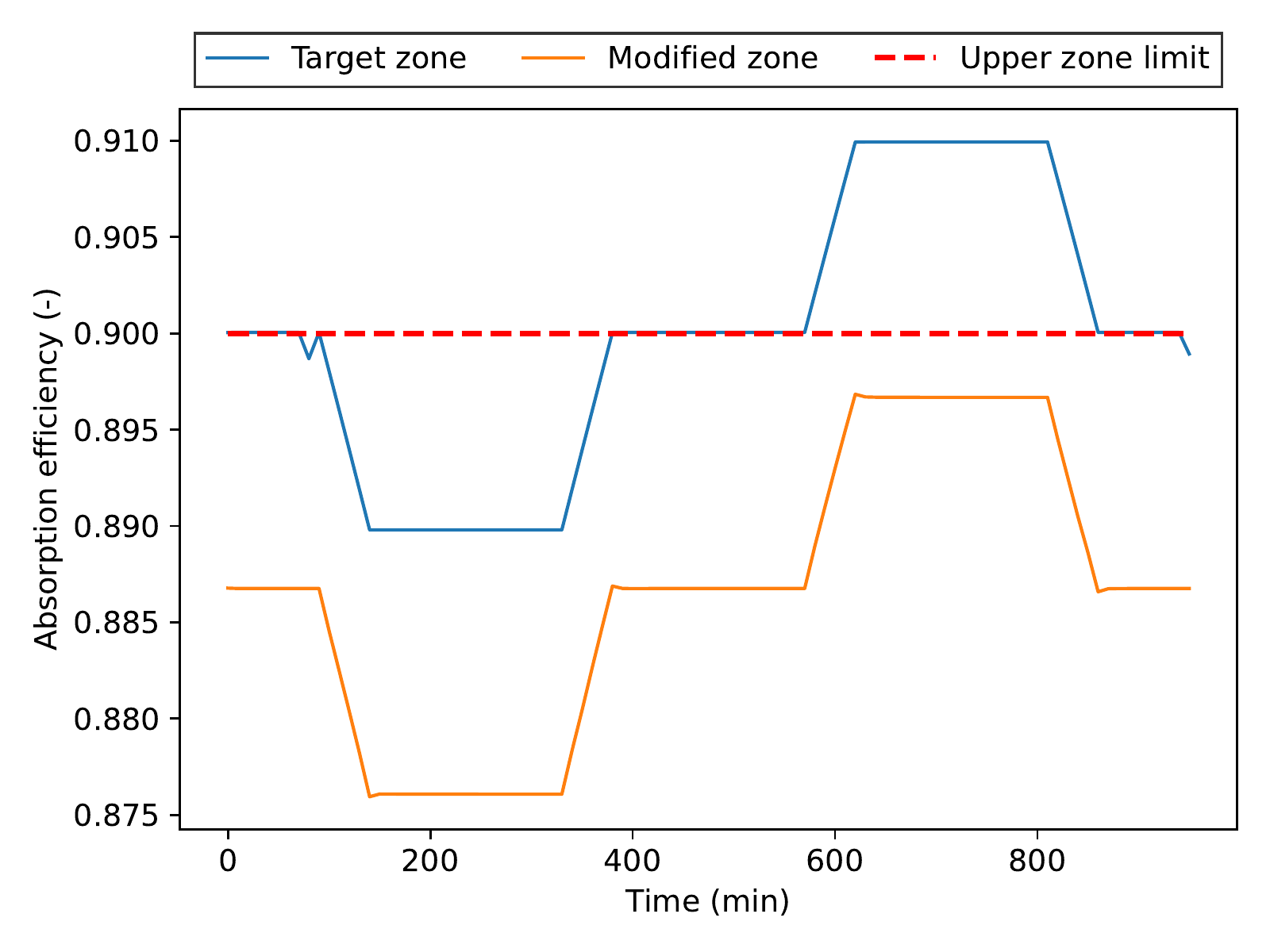}    
    \caption{Trajectories of the absorption efficiency for the absorption column under the operation of the zone EMPC control algorithm tracking the original target zone (blue) and modified zone (orange) for the time-varying flue gas scenario} 
    \label{fig:tvf_abs}
  \end{center}
\end{figure}

\begin{figure}[tp!]
  \begin{center}
    \includegraphics[width=0.6\textwidth]{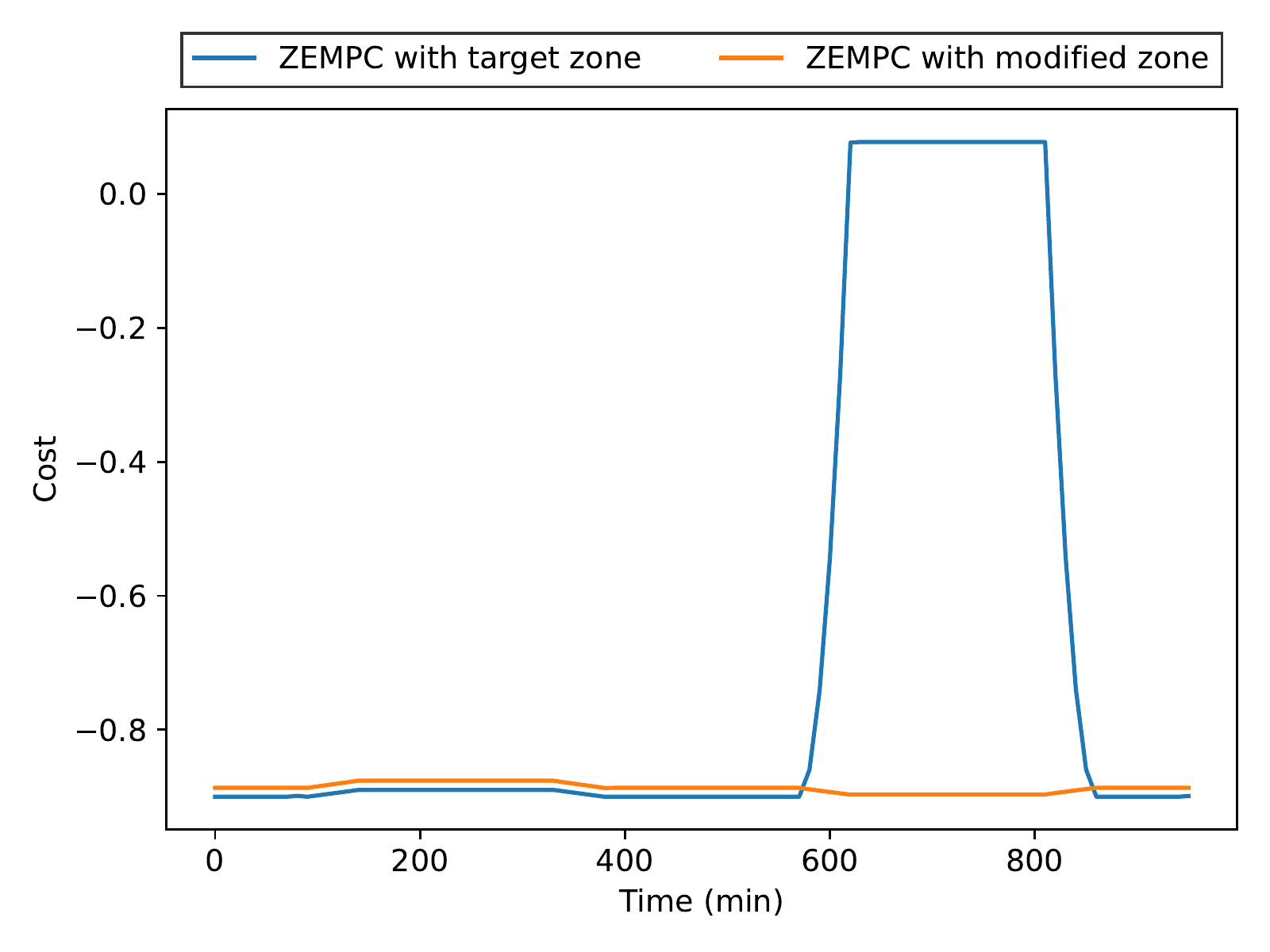}    
    \caption{Trajectories of the stage cost for the absorption column under the operation of the zone EMPC control algorithm tracking the original target zone (blue) and modified zone (orange) for the time-varying flue gas flow rate} 
    \label{fig:tvf_cost}
  \end{center}
\end{figure}

We compare the performance of the EMPC with target zone tracking (NZEMPC) to that of the EMPC with modified zone tracking (RZEMPC) under a time-varying flue gas flow rate setting. This was achieved by varying the flue gas flow rate using the disturbance shown in Figure~\ref{fig:tvf_flow}. 
\begin{table}[tp!]
  \begin{center}
  \caption{Comparison of the average cost of NZEMPC and RZEMPC under time-varying flue gas flow rate (smaller is better) }\label{tb:tvf_result_comparison}
  \vspace{2mm}
    \begin{tabular}{lccc}
        \hline
    Zone & Ramping up cost & Ramping down cost & Average overall cost \\  
        \hline
    Target zone & $-0.8948$ & $-0.4525$ & $ -0.6690$ \\ 
    Modified zone  & $-0.8813$ & $-0.8916$ & $-0.8866$ \\ 
        \hline
    \end{tabular}
  \end{center}
\end{table}
The disturbance to the flue gas mimics typical ramp up and ramp down behaviour of a power plant. The average costs of the operation of the absorption column under the two controllers is presented in Table~\ref{tb:tvf_result_comparison}. As can be seen the EMPC with modified zone tracking yields a better overall cost on average than that of the EMPC with target zone tracking. The reason for the poor performance in the NZEMPC is the same as explained in the earlier section. The economic objective drives the absorption efficiency to the boundary of the target zone which leads to a target zone violation once the disturbance is present. The RZEMPC on the other hand causes the system to operate away from the boundary thus making room for the effects of the disturbance. This ensures that the absorption efficiency stays within the target zone at all times which leads to a better cost on average. The absorption efficiency trajectory can be seen in Figure~\ref{fig:tvf_abs}. It can be seen that in both cases, the absorption efficiency decreases during the ramp up and increases during the ramp down. This is because when the flue gas flow rate increase, the amount of CO$_2$ entering the column also increase. The controllers still try to capture the same amount of CO$_2$ since the process model used in the controller uses the nominal flue gas flow rate. A careful look at the cost trajectories in Figure~\ref{fig:tvf_cost} and as shown in Table~\ref{tb:tvf_result_comparison} shows that during the ramping up phase, the NZEMPC yields a better cost than that of the RZEMPC. However, the NZEMPC performs poorly during the ramping down phase. The RZEMPC on the other hand ensures a fairly constant cost throughout the operation. This shows the benefits of modifying the target zone to ensure finite time zone tracking.

\section{Concluding remarks}

In this work, a control problem which typically arises in the control of the absorption unit in a post-combustion CO$_2$ capture plant is addressed using an EMPC with zone tracking formulation. This helps to avoid the problems of solvent overcirculation and flooding in the column during operation. To ensure that finite-time zone tracking objective is achieved in the presence of modelling uncertainty, the target zone to be tracked is modified. The proposed zone modification algorithm makes use of ellipsoidal control invariant set and a back-off strategy which is scalable for systems with large number of states such as the absorption column. The use of the control invariant set as the zone ensures that the zone can be tracked since there is no guarantee that the original target zone is control invariant. The simulation example demonstrates the efficacy of the proposed EMPC algorithm with zone tracking as a effective control strategy for the absorption column of a typical post-combustion CO$_2$ capture plant.

In the future, it would be interesting to apply the zone EMPC algorithm to the entire PCC plant. Another direction is the extension of the zone modification algorithm to other general nonlinear systems.

\section{Acknowledgement}
This work is supported in part by the Natural Sciences and Engineering Research Council of Canada.


\end{document}